**TITLE**

Direct homologous dsDNA-dsDNA pairing: how, where and why?

**AUTHORS**

Alexey K. Mazur[1,2,3], Tinh-Suong Nguyen[2], Eugene Gladyshev[2]*

**AFFILIATION**

[1] CNRS, Université de Paris, UPR 9080, Laboratoire de Biochimie Théorique, 13 rue Pierre et Marie Curie, F-75005, Paris, France

[2] Group Fungal Epigenomics, Department of Mycology, Institut Pasteur, Paris 75015, France

[3] Institut de Biologie Physico-Chimique-Fondation Edmond de Rothschild, PSL Research University, Paris, France

*Corresponding address: eugene.gladyshev@gmail.com

**TYPE**

Perspective

**DECLARATIONS OF INTEREST**: none

**ABSTRACT**

The ability of homologous chromosomes (or selected chromosomal loci) to pair specifically in the apparent absence of DNA breakage and recombination represents a prominent feature of eukaryotic biology. The mechanism of homology recognition at the basis of such recombination-independent pairing has remained elusive. A number of studies have supported the idea that sequence homology can be sensed between intact DNA double helices *in vivo*. In particular, recent analyses of the two silencing phenomena in fungi, known as "repeat-induced point mutation" (RIP) and "meiotic silencing by unpaired DNA" (MSUD), have provided genetic evidence for the existence of the direct homologous dsDNA-dsDNA pairing. Both RIP and MSUD likely rely on the same search strategy, by which dsDNA segments are matched as arrays of interspersed base-pair triplets. This process is general and very efficient, yet it proceeds normally without the RecA/Rad51/Dmc1 proteins. Further studies of RIP and MSUD may yield surprising insights into the function of DNA in the cell.

**I. Homologous dsDNA-dsDNA pairing: how?**

The *in vivo* occurrence of direct homologous dsDNA-dsDNA pairing remains hypothetical. Two atomic models of sequence-specific interactions between long intact dsDNA molecules have been formulated and validated by molecular-dynamics analysis[1,2] (Fig. 1). In both models, homologous dsDNAs pair at periodically spaced sites by forming four-stranded structures. The models differ substantially with respect to the organization of such paired structures (Fig. 1). According to the first model, proposed by Seeman and colleagues[1], each pairing site corresponds to the paranemic crossover (PX) motif, created by the reciprocal exchange of both strands between the co-aligned DNA duplexes (Fig. 1, "PX-DNA"). To engage in this type of pairing, B-DNA duplexes must melt to break the existing intra-helical WC bonds and then establish new inter-helical WC bonds by wrapping around each other. Because of this inter-wrapping configuration, the unit homology length must be larger or equal to approximately half-turn of the original B-DNA duplex (5-6 base-pairs). Previously, the PX-DNA motif was used as a building block in DNA nanotechnology, and its formation was reported to occur spontaneously between DNA repeats in the context of a negatively supercoiled plasmid[1].

The second model (Fig. 1, "PB-DNA") is based on the geometric property of the standard WC base-pairs to be self-complementary at their major-groove surfaces[3,4]. As a result, identical base-pairs can form planar quartets (AT/AT and GC/GC), each maintained by two supplementary hydrogen bonds[3,4]. This long-known principle of self-complementarity provides a theoretical basis for the homologous pairing of long dsDNAs[5]. Recent analysis by all-atom molecular dynamics and quantum mechanics calculations has suggested that the high flexibility of B-DNA double helices may indeed permit direct homologous contacts between the major groove surfaces, but only along short stretches of 3-4 base-pairs[2]. The predicted contact site represents a quadruplex formed by two deformed B-DNAs (Fig. 1, "PB-DNA"). Unfavorable energy of deformations is compensated by ion-mediated and electron-polarization electrostatic interactions established between identical base-pairs[2]. This energy trade-off is expected to favor specific pairing of homologous dsDNAs[2]. Overall, the two double helices are juxtaposed in a locally parallel but globally crossed orientation (Fig. 1, "PB-DNA"). The predicted shape of the dsDNA-dsDNA crossing is expected to constrain the spacing of the quadruplex contacts along the helices. Most notably, these contacts should be separated by a few helical turns and coherent with the helical rotation[2].

**II. Homologous dsDNA-dsDNA pairing: where?**

The ability of homologous chromosomes to preferentially associate with each other in the apparent absence of DNA breakage and recombination is well documented[6]. Mammals feature a particularly rich repertoire of such recombination-independent phenomena[6,7]. For example, homologous chromosomes still pair transiently during early meiosis in the absence of the programmed DNA breaks in mice[8]. Such recombination-independent meiotic pairing was also observed in other model organisms, including the fruit fly *D. melanogaster*[9], the round worm *C. elegans*[10], and the fission yeast *S. pombe*[11]. In early mammalian development, some homologous loci engage in extensive pairing, presumably to establish appropriate patterns of gene expression[12]. Transient association of X chromosomes prior to random X-chromosome inactivation provides arguably the most well-known case of such developmentally-regulated pairing[13]. Recombination-independent pairing in somatic cells is by no means restricted to mammals, and it has also been described in many other organisms, including the fruit fly (discussed below) and budding yeast *S. cerevisiae*[14,15,16].

Mammalian genomes also contain large amounts of repetitive ("self-homologous") DNA normally silenced in the form of constitutive heterochromatin[17]. Importantly, in mammals, the initiation of heterochromatin assembly on tandemly repeated DNA does not require RNA interference[18,19]. Pathological misregulation of this process is associated with several types of cancer[20,21] and other disease, such as Type I Facioscapulohumeral muscular dystrophy[22]. The nature of the mechanism(s) responsible for the identification of repetitive DNA sequences remains largely unknown, but has been proposed to involve pair-wise interactions between repeat units[23,24].

The classical example of recombination-independent somatic pairing was discovered in *D. melanogaster* and other Diptera insects more than a century ago[25]. In these animals, homologous chromosomes are associated in the majority of non-dividing nuclei throughout the entire life of the organism[25]. Yet this pairing state is dynamic and involves both forward (association) and reverse (dissociation) reactions[26]. Importantly, in *D. melanogaster*, somatic pairing is critical for ensuring proper transvection, a genetic phenomenon in which the two allelic copies of a gene comprise one unit of expression[25]. Recent genome-wide studies using haplotype-resolved Hi-C[27,28] and super-resolution microscopy[29] provided two additional insights into the nature of this process. First, they have demonstrated that homologous chromosomes are indeed paired throughout their entire lengths, with regions of tight pairing interrupted by regions of loose pairing. Second, these studies have also found that in the tightly paired regions, the interacting allelic segments cannot be distinguished at the highest achievable resolution, suggesting that homologous DNA duplexes may be intimately (perhaps directly) associated.

The strongest evidence in support of the existence of the direct homologous dsDNA-dsDNA pairing *in vivo* has been provided by two gene silencing processes in the filamentous fungus *Neurospora crassa*. Both processes appear to recognize DNA sequence homology by a cardinally new general mechanism that does not require eukaryotic RecA-like recombinases (Rad51 or Dmc1) and, instead, apparently matches DNA double helices as arrays of interspersed base-pair triplets[30,31]. These processes are known as "repeat-induced point mutation" (RIP) and "meiotic silencing by unpaired DNA" (MSUD).

**Repeat-induced point mutation (RIP)**
RIP is a genome-defense process that detects and mutates gene-sized DNA repeats largely irrespective of their origin, transcriptional capacity, as well as relative or absolute positions in the genome[32,33]. RIP is activated in the haploid nuclei that continue to divide by mitosis in preparation for karyogamy, followed by meiosis[32,33]. These premeiotic nuclei are found in specialized ascogenous hyphae inside a complex mating structure and therefore comprise a dedicated generative lineage that may be considered as a germ-line. RIP was discovered in *N. crassa* nearly 30 years ago[33] and subsequently demonstrated experimentally in many other filamentous fungi[34,35]. More broadly, signatures of RIP mutation can be found by bioinformatic analysis in nearly all filamentous fungi[34,35,36]. A process analogous to RIP and known as "methylation induced premeiotically" (MIP) also occurs in another filamentous fungus *Ascobolus immersus*[37,38]. Similarly to RIP, MIP detects gene-size DNA duplications but then silences them by cytosine methylation instead of C-to-T mutation[37,38].

*In N. crassa*, RIP mutation is mediated by the two pathways[24]. The first pathway requires a putative C5-cytosine methyltransferase RID[39]. An ortholog of RID is also required for *de novo* cytosine methylation during MIP[37,38]. RID-like proteins comprise an ancient evolutionary clade, yet their current phylogenetic distribution is restricted to filamentous fungi[40]. The second RIP pathway requires DIM-2, the only other C5-cytosine methyltransferase in *N. crassa*[41]. The two pathways show clear substrate preferences: the RID-dependent RIP targets DNA repeats *per se*, whereas the DIM-2-dependent RIP is localized preferentially in the intervening regions between closely-positioned direct repeats[24,42]. In *N. crassa*, the activity of DIM-2 is normally controlled by DIM-5, a conserved SUV39 lysine methyltransferase that catalyzes trimethylation of histone H3 lysine-9 residues (H3K9me3)[41]. Orthologs of DIM-5 are present in other fungi, animals (including humans), plants and unicellular eukaryotes, where they mediate the formation of H3K9me2/3 at tandemly repeated DNA[43]. In *N. crassa*, DIM-2 is normally guided to H3K9me3 by HP1[41]. Importantly, HP1 and DIM-5 are both required for the DIM-2-dependent RIP[24], raising a possibility that the SUV39 proteins can be recruited to repetitive DNA by a mechanism that involves homologous dsDNA-dsDNA pairing[24].

Systematic testing of many different homology patterns for their ability to promote RIP revealed the existence of a cardinally new mechanism by which DNA molecules were evaluated for sequence identity[30]. Specifically, it was discovered that a series of short microhomologies (as short 3 base-pairs) could still promote RIP[30]. To do so effectively, homologous units had to be spaced regularly with a periodicity of 11 or 12 base-pairs, and over the total length of several hundred base-pairs[30]. All attempts to interfere with this characteristic pattern abrogated RIP[30]. These genetic results suggested that homology was likely detected directly between long co-aligned DNA double helices. Notably, the process of homology recognition could be decoupled from mutation at the level of individual base-pairs[44], further supporting the idea that a relatively long DNA segment (of several hundred base-pairs) must be defined as homologous at once.

**Meiotic silencing by unpaired DNA (MSUD)**
In addition to RIP, *N. crassa* features another genome-defense process known as "meiotic silencing by unpaired

DNA" (MSUD)[45,46]. The hallmark of MSUD is its remarkable ability to efficiently detect non-homologous DNA sequences present at the allelic positions on a pair of homologous chromosomes[46] (Fig. 2A). Once detected, such "unpairable" loci undergo transient post-transcriptional silencing by RNA interference[46]. Because the last round of premeiotic DNA replication precedes karyogamy in *N. crassa* and other related fungi, MSUD needs to operate on the already replicated chromosomes during a time interval of only a few hours, after karyogamy but before the onset the recombination-dependent pairing stage[45,47,48].

The combined effort of several labs over nearly twenty years has identified a large number of factors involved in MSUD[46,49]. These factors were isolated either by a candidate gene approach (by testing the genes with known roles in the somatic RNA interference pathway known as "quelling") or through a screen for gene deletions that would work as dominant suppressors of MSUD[49]. This screen was based on the idea that a gene-deletion allele could induce MSUD-mediated silencing of the corresponding wildtype allele, leading to a partial attenuation of silencing of the unpaired reporter[49].

The capacity of MSUD to recognize non-homologous DNAs as short as 1 kbp, irrespective of their origin and transcriptional capacity, suggests that MSUD involves a general and efficient homology search[46]. Recent work has found that MSUD likely shares its mechanism of homology recognition with RIP[31]. First, similarly to RIP, MSUD can sense a long series of homologous triplets interspersed along the allelic dsDNAs[31]. Second, MSUD also proceeds normally in the absence of MEI-3 (the only RecA-like recombinase in *N. crassa*) and Spo11 (type II DNA topoisomerase-like protein that is responsible for generating programmed DNA breaks in meiosis[47]).

In summary, the widespread occurrence of recombination-independent homologous pairing contrasts with our limited understanding of its mechanistic basis. Recent studies in the model fungus *N. crassa* have revealed the existence of a new mechanism by which sequence information can be sensed between chromosomes or selected chromosomal loci. This mechanism appears to be cardinally different from all the known canonical forms of homology recognition in its ability to match segments of chromosomal DNA as arrays of interspersed base-pair triplets (Fig. 2B). The involvement on this same mechanism in both RIP and MSUD hints at a possibility that it may encompass a general solution to the problem of recombination-independent homology recognition.

**III. Homologous dsDNA-dsDNA pairing: why?**
In principle, the ability of homologous dsDNAs to engage in stable and specific pairing interactions could be used *in vivo* for several purposes. Three broad categories are provided as examples below.

**1. A general strategy for co-localizing and co-orienting homologous chromosomal regions**
It is conceivable that each instance of recombination-independent homologous pairing relies on a specialized mechanism. Such mechanisms, for example, may involve indirect pairing by sequence-specific proteins or non-coding RNAs. A classical example is provided by the juxtaposition of the two *lacO* operators by LacI protein, with the concomitant formation of a short DNA loop[50]. More recently, the roles of various proteins as mediators of homologous pairing has been revisited by several studies. For example, Ishiguro, Watanabe and colleagues, focusing on early meiosis in mouse spermatocytes, proposed a model in which homologous chromosomes exhibit specific longitudinal patterns of meiotic cohesins that uniquely define these chromosomes for pairing[51]. In this situation, DNA sequence is read and recognized indirectly as a chromosome-long protein "bar-code" (ref. 51). In another study, Kim, Shendure and colleagues identified a strong instance of inter-allelic pairing at the *HAS1pr-TDA1pr* locus during the stationary growth phase in budding yeast and have shown that it requires a combination of the three transcription factors (TFs), Leu3, Sdd4, and Rgt1[52].

While the role of several DNA-binding proteins in co-localizing homologous loci has been documented, it is unclear if the protein-mediated co-localization can support pairing of larger chromosomal regions or the entire chromosomes. The widespread occurrence of the latter, observed in a wide range of organisms from mammals to bacteria[53], suggests that a conserved mechanism could be involved. In principle, such a mechanism may be mediated by RNAs. For example, Ding, Hiraoka and colleagues discovered the ability of one abundant meiotic-specific non-coding RNA, transcribed from the *sme2* locus, to drive pairing of the two *sme2* loci in the fission yeast *S. pombe*[54]. Presumably, other unidentified RNA species, transcribed from many additional loci, can promote recombination-independent pairing of whole chromosomes. This idea is naturally appealing because RNAs can associate with double-stranded DNA in a sequence-specific manner, either by forming triplexes or R-loops[55]. This general property of RNA can potentially support a universal recombination-independent pairing mechanism. Yet it is important to note that the *sme2* pairing itself represents a highly specialized situation, as it relies on the dedicated RNA species that accumulates in very large amounts at the site of its transcription, the *sme2* locus, and acts as a decoy to sequester the Mmi1 protein, which normally inhibits meiosis by promoting the elimination of meiosis-specific transcripts[56]. In the light of these considerations, the general role of RNAs as mediators of recombination-independent pairing remains hypothetical, at least for now.

In the absence of a definitive mechanism, the possibility of the sequence-specific dsDNA-dsDNA interactions at the basis of recombination-independent homologous pairing represents an intriguing alternative. One advantage of the direct dsDNA-dsDNA pairing is the lack of the requirement for numerous adaptor molecules, which should reduce its entropic cost. Another advantage is its high precision, which can only be approximated by the indirect mechanisms.

The direct dsDNA-dsDNA pairing may still require additional factors, for example to remodel chromatin and make DNAs more accessible, or to promote the pairing/unpairing reaction itself. In principle, the direct dsDNA-dsDNA pairing may synergize with indirect mechanisms. For example, the indirect mechanism may be used to reduce the complexity of the search space that needs to be explored by the direct mechanism. On the other hand, RIP is able to routinely detect *any* two gene-sized DNA repeats that could be present at any arbitrary genomic positions, strongly suggesting that RIP involves a comprehensive "genome-by-genome" homology search. To be feasible at the relevant time scales, this search process must be extremely efficient.

## 2. Paired dsDNAs as a specific substrate for certain biochemical or signaling pathways

In addition to serving as a general basis for co-localizing and co-orienting homologous chromosomal regions, the direct dsDNA-dsDNA pairing can provide a specific substrate for biochemical or signaling pathways. Two potential examples are noted below.

### Recognition of the parallel dsDNAs by the catalytic domain of Dnmt3a

In mammals, chromosomal regions containing large amounts of tandemly repeated DNA (such as centromeres and pericentromeres) are normally associated with high levels of cytosine methylation established by the two *de novo* C5-cytosine methyltransferases, Dnmt3a and Dnmt3b[57,58]. How these proteins are recruited specifically to repetitive DNA remains incompletely understood. Jurkowska, Jeltsch and colleagues have reported that (i) the oligomerized catalytic domain of Dnmt3a can bridge two long parallel dsDNA molecules, and (ii) the ability to oligomerize is required for the recruitment of Dnmt3a to repetitive DNA *in vivo*[59] (also see ref. 58 for the earlier discussion of this idea). Given the recently proposed role of the direct homologous dsDNA-dsDNA pairing in guiding the initiation of transcriptional silencing[24], these results revive an idea[60] that cytosine methylation can be induced in response to homologous dsDNA-dsDNA pairing.

### Recognition of the PX-joined parallel dsDNAs by *E. coli* DNA polymerase I

In an attempt to identify proteins that specifically recognize PX-DNA (above), Gao, Seeman and colleagues have found that DNA polymerase I from the bacterium *E. coli* has surprisingly high affinity for the synthetic complex containing two dsDNAs joined by the PX motif[61]. Docking the structure of the Klenow fragment of Polymerase I onto the predicted structure of PX-DNA produced a complex in which two polymerase molecules were bound (perhaps as a dimer) to the paired dsDNAs. While the significance of this finding cannot be known with certainty at the present moment, the authors have noted that the ability of DNA polymerase I to bind paired dsDNAs may be involved in DNA repair and recombination[61].

## 3. Homologous dsDNA-dsDNA pairing as a general mechanism to generate DNA supercoiling

The efficient recognition of interspersed homologies during RIP suggests that the two paired double helices must maintain their twist rigidity[30]. The optimal sequence periods appear to be close to, but slightly different from, that of helical rotation in B-DNA suggesting that the paired DNA molecules are engaged in interactions that affect their helical twisting. The PB-DNA model can account for these results and also predicts that pairing can be stabilized if the two double helices can fold in a plectoneme[2]. Taken together, the above observations point at a possible link between homologous dsDNA-dsDNA pairing and DNA supercoiling. On one hand, the homology recognition can be facilitated in the presence of type II topoisomerases and, consequently, regulated by them. On the other hand, small concerted twisting of long dsDNA stretches within a supercoiled segment can produce significant conformational changes on larger scales and induce DNA bending, looping, etc. A putative manifestation of this effect has been found in the striking difference between the patterns of RIP mutations for the pairs of closely-positioned repeats present in either "head-to-tail" or "head-to-head" orientations[42].

Overall, it is possible that the direct dsDNA-dsDNA pairing may be used as an additional general mechanism of generating an excess of DNA supercoiling of both signs, depending upon the exact configuration of the paired DNAs as well as the nature of the recruited factors (such as histone cores, HMG proteins, or topoisomerases). DNA supercoiling generated by pairing of multiple closely positioned repeats may play critical and conserved roles in specifying unique properties of chromosomal regions with large amounts of repetitive DNA[24,42].

In summary, the direct homologous dsDNA-dsDNA pairing, if it indeed takes place *in vivo*, may play many different roles in a number of essential cytogenetic processes. Further studies of RIP and MSUD in *N. crassa* from a biophysical perspective may lead to the discovery of the recombination-independent mechanism of DNA homology search and recognition.

**FIGURE LEGENDS**

**Figure 1. Two alternative models of the direct homologous dsDNA-dsDNA pairing**

Both models were first assembled by manually docking and re-linking dsDNA segments using custom software. The model of PX-DNA ("Paranemic crossover DNA", ref. 1) was built from two canonical B-DNAs containing alternating GC sequences (ref. 62). The model of PB-DNA ("Paired B-DNA", ref. 2) was built by extending a 15-mer paired complex (corresponding to a molecular-dynamics snapshot from ref. 2) with canonical B-DNAs in the four directions. Both models were then subjected to vacuum energy minimization (with original helices harmonically restrained to initial conformations).

**Figure 2. Repeat-induced point mutation (RIP) and meiotic silencing by unpaired DNA (MSUD)**

(A) RIP is activated in the haploid premeiotic nuclei that continue to divide by mitosis. RIP detects and mutates (by numerous C-to-T transitions) gene-sized DNA repeats irrespective of their origin, sequence composition, transcriptional or coding capacity and positions in the genome. Subsequently, MSUD takes place in the diploid (meiotic) nucleus shortly after karyogamy but before the onset of recombination-mediated pairing. MSUD identifies regions of no-homology (~1 kbp or longer) present at allelic chromosomal positions. Once detected, these "unpairable" regions undergo transient post-transcriptional silencing by RNA interference.

(B) RIP and MSUD appear to share the same recombination-independent mechanism of homology recognition that cross-matches segments of chromosomal DNA as arrays of interspersed base-pair triplets. In doing so, this mechanism is cardinally different from other known sequence-specific interactions involving aligned DNA and RNA molecules.

**ACKNOWLEDGEMENT**

The work was supported by CNRS (A.K.M), the ANR "Laboratoires d'excellence" programs 11-LABX-0011 "DYNAMO" (A.K.M) and 10-LABX-0062 "IBEID" (E.G.), Institut Pasteur (E.G.), and Fondation pour la Recherche Médicale grant AJE20180539525 (T.N. & E.G.).

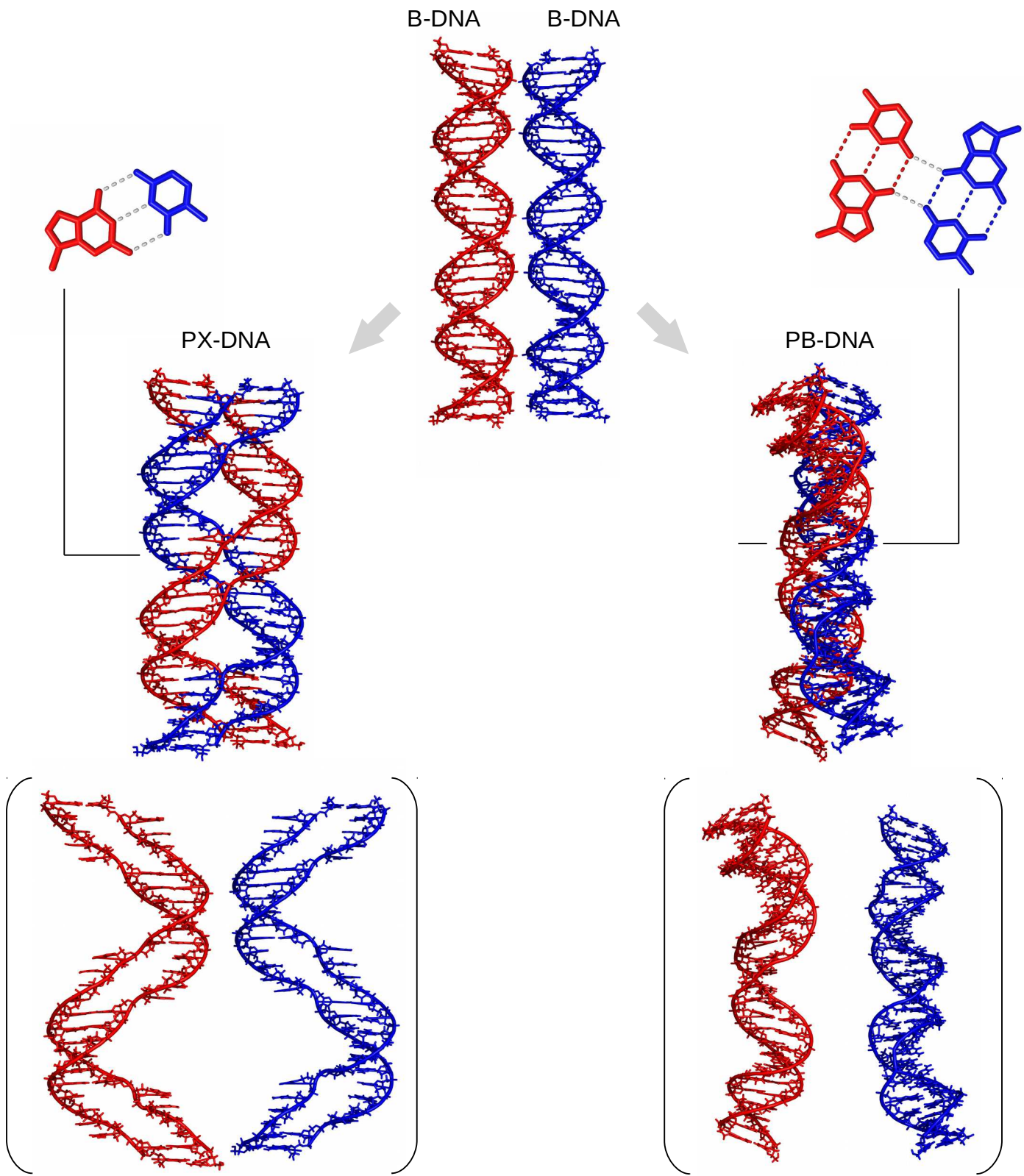

**Figure 1. Two alternative models of the direct homologous dsDNA-dsDNA pairing**
Both models were first assembled by manually docking and re-linking dsDNA segments using custom software. The model of PX-DNA ("Paranemic crossover DNA", ref. 1) was built from two canonical B-DNAs containing alternating GC sequences (ref. 62). The model of PB-DNA ("Paired B-DNA", ref. 2) was built by extending a 15-mer paired complex (corresponding to a molecular-dynamics snapshot from ref. 2) with canonical B-DNAs in the four directions. Both models were then subjected to vacuum energy minimization (with original helices harmonically restrained to initial conformations).

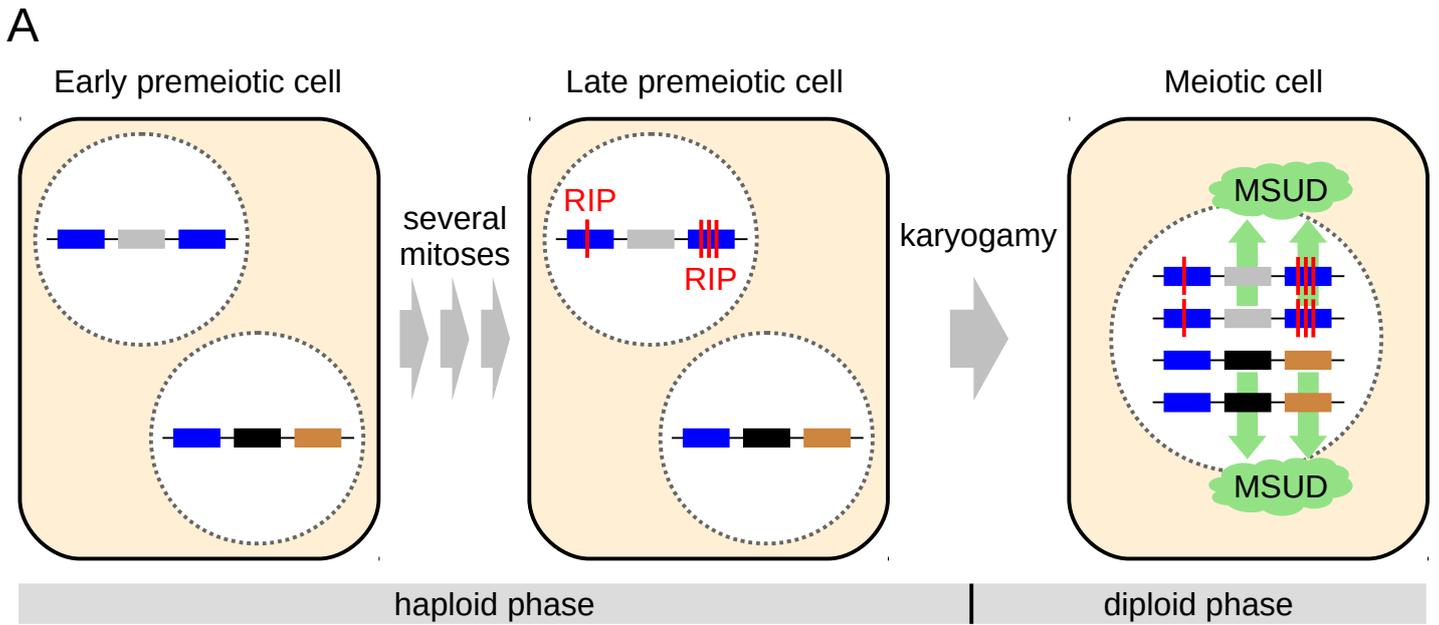

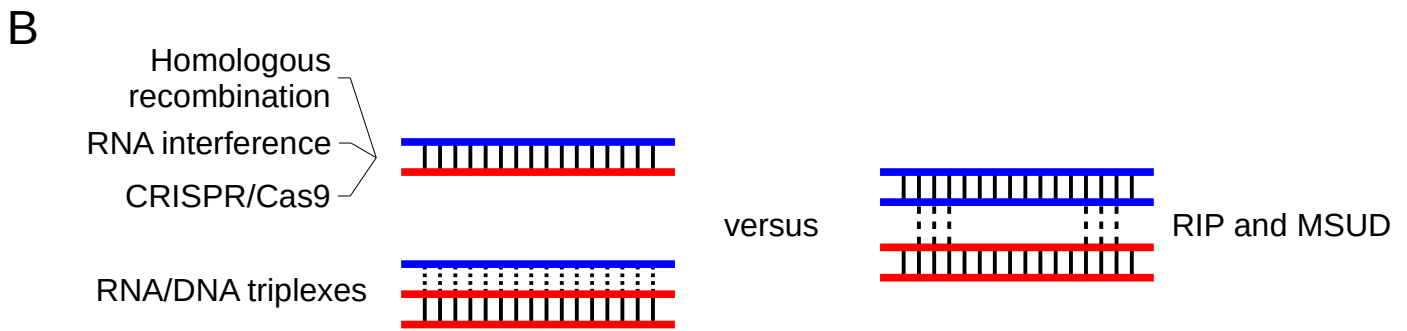

**Figure 2. Repeat-induced point mutation (RIP) and meiotic silencing by unpaired DNA (MSUD)**

(A) RIP is activated in the haploid premeiotic nuclei that continue to divide by mitosis. RIP detects and mutates (by numerous C-to-T transitions) gene-sized DNA repeats irrespective of their origin, sequence composition, transcriptional or coding capacity and positions in the genome. Subsequently, MSUD takes place in the diploid (meiotic) nucleus shortly after karyogamy but before the onset of recombination-mediated pairing. MSUD identifies regions of no-homology (~1 kbp or longer) present at allelic chromosomal positions. Once detected, these "unpairable" regions undergo transient post-transcriptional silencing by RNA interference.

(B) RIP and MSUD appear to share the same recombination-independent mechanism of homology recognition that cross-matches segments of chromosomal DNA as arrays of interspersed base-pair triplets. In doing so, this mechanism is cardinally different from other known sequence-specific interactions involving aligned DNA and RNA molecules.